\documentclass[acmtog,nonacm, nonaddress]{acmart}
\acmSubmissionID{2503}

\usepackage{booktabs} 

\makeatletter
\let\@authorsaddresses\@empty
\makeatother

\usepackage[ruled]{algorithm2e} 

\SetAlFnt{\small}
\SetAlCapFnt{\small}
\SetAlCapNameFnt{\small}
\SetAlCapHSkip{0pt}

\acmJournal{TOG}

\usepackage{xspace}

\makeatletter
\DeclareRobustCommand\onedot{\futurelet\@let@token\@onedot}
\def\@onedot{\ifx\@let@token.\else.\null\fi\xspace}

\def\eg{\emph{e.g}\onedot} 
\def\ie{\emph{i.e}\onedot} 
 
\def\etc{\emph{etc}\onedot}

\makeatother

\begin{document}
\title{FairyGen: Storied Cartoon Video from a Single Child-Drawn Character}

\author{Jiayi Zheng}
\author{Xiaodong Cun}
\authornote{Corresponding Author}
\affiliation{
    \institution{GVC Lab, Great Bay University}
    \city{Guangdong}
    \country{China}
    }


\settopmatter{printacmref=false}

\begin{abstract}
We propose \textbf{FairyGen}, an automatic system for generating story-driven cartoon videos from a single child’s drawing, while faithfully preserving its unique artistic style. Unlike previous storytelling methods that primarily focus on character consistency and basic motion, FairyGen explicitly disentangles character modeling from stylized background generation and incorporates cinematic shot design to support expressive and coherent storytelling.
Given a single character sketch, we first employ a Multimodal Large Language Model (MLLM) to generate a structured storyboard with shot-level descriptions that specify environment settings, character actions, and camera perspectives. To ensure visual consistency, we introduce a \textit{style propagation adapter} that captures the character’s visual style and applies it to the background, faithfully retaining the character’s full visual identity while synthesizing style-consistent scenes. A \textit{shot design module} further enhances visual diversity and cinematic quality through frame cropping and multi-view synthesis based on the storyboard.
To animate the story, we reconstruct a 3D proxy of the character to derive physically plausible motion sequences, which are then used to fine-tune an MMDiT-based image-to-video diffusion model. We further propose a two-stage motion customization adapter: the first stage learns appearance features from temporally unordered frames, disentangling identity from motion; the second stage models temporal dynamics using a \textit{timestep-shift strategy} with frozen identity weights.
Once trained, FairyGen directly renders diverse and coherent video scenes aligned with the storyboard. Extensive experiments demonstrate that our system produces animations that are stylistically faithful, narratively structured, and rich in smooth, natural motion, highlighting its potential for personalized and engaging story animation.
\end{abstract}
\begin{teaserfigure}
\centering
  \includegraphics[width=1\textwidth]{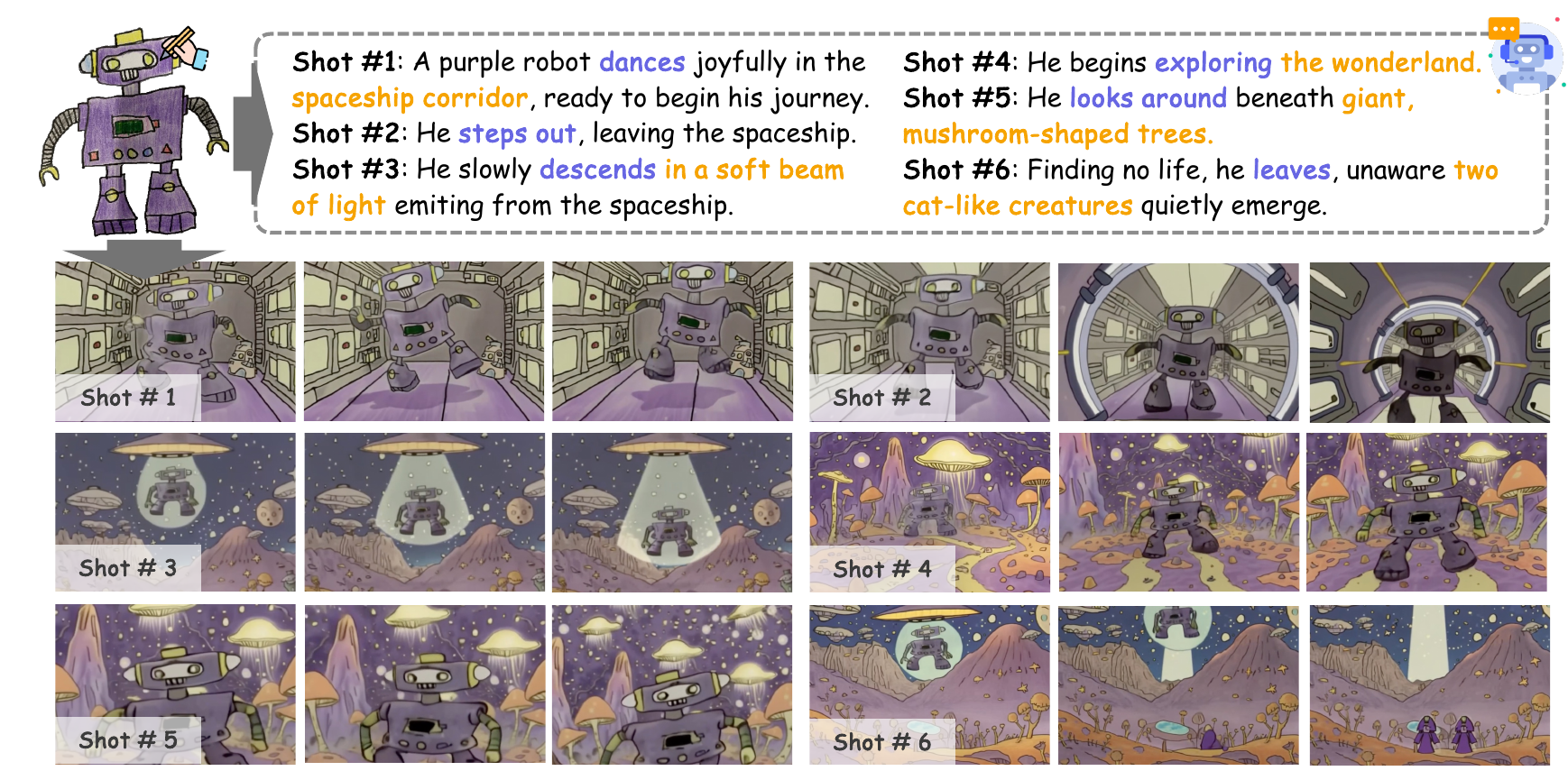}
  \vspace{-2em}
  \caption{We present FairyGen, a visual story generation framework to generate multi-shot cartoon videos from a single child-drawn character with consistent style and motion between the foreground and the background. Project page: {\color{red}\url{https://jayleejia.github.io/FairyGen/}}. }
  \label{fig:teaser}
\end{teaserfigure} 

\maketitle
\newcommand{\todo}[1]{\textcolor{red}{TODO #1}}
\newcommand{\tocite}{\textcolor{red}{TO Cite}}
\newcommand{\modelname}{LVDM\xspace}

\section{Introduction}
Children often express vivid imagination through abstract, stylized drawings featuring simple cartoon characters and expressive visual elements. Despite their lack of photorealistic detail, these illustrations convey unique artistic styles and emotional intent. Transforming such drawings into coherent animated stories bridges youthful creativity with expressive storytelling, offering promising applications in education, digital art therapy, personalized content creation, and interactive entertainment.
Story visualization has become a major research area in computer graphics, with recent advances in generative video model continue to expand its horizons.
Prior works such as  StoryGAN~\cite{storygan}, AR-LDM~\cite{ar-ldm} and Make-A-Story~\cite{make-a-story} improve visual fidelity and semantic coherence, but are limited by style diversity and training data constraints.
More recent LLM-driven pipelines like TaleCrafter~\cite{talecrafter}, DreamStory~\cite{dreamstory}, and Animate-A-Story~\cite{animate-a-story} introduce modular task decomposition for better controllability but often suffer from character inconsistency, fragmented narratives, and weak motion quality.
Diffusion-based video models such as MEVG~\cite{mevg}, MovieDreamer~\cite{moviedreamer}, and Vlogger~\cite{vlogger} improve temporal consistency and narrative flow, yet still struggle with cross-shot character consistency, artistic style preservation, and complex motion synthesis—primarily due to reliance on real-world data priors.
These limitations become more pronounced when dealing with abstract hand-drawn characters, especially in single-example scenarios where the input style diverges significantly from training data. 
To address this, we consider cartoon story generation with a decoupled design that explicitly separates the foreground character from background synthesis. Using 3D reconstruction that adheres to physical constraints, we preserve the character identity while enabling plausible motion generation. In coordination, background generation is treated as a style adaptation process that propagates the character’s visual style to scene elements \cite{brooks2022hallucinating, actanywhere, kulal2023putting}. This decoupled design further extends to video generation, where first learns the motion priors from 3D-derived sequences, and then animate the full scene using a large pre-trained video diffusion model. This pipeline naturally preserves character consistency, supports complex motion, and enables cinematic storytelling.
Building on these insights, we present FairyGen, a novel framework for generating animated story videos from a single hand-drawn character. FairyGen produces expressive motion, stylistically aligned backgrounds, and cinematic compositions without requiring additional training data.
Specifically, given a hand-drawn character, we first generate a structured storyboard that describes actions, settings, and camera compositions using an MLLM. 
To ensure visual consistency, we propose a style propagation adapter that preserves the character’s  full visual identity while learning its stylistic traits, which are then propagated to the background using a pre-trained inpainting diffusion model.
Next, to animate the story, we reconstruct a 3D proxy of the character and derive physically plausible motion sequences, which are then used to finetune an MMDiT-based image-to-video diffusion model~\cite{wanvideo}. For robust motion synthesis, we introduce a two-stage motion customization adapter: the first stage learns spatial features from temporally shuffled frames to remove temporal bias, and the second stage captures dynamics via a novel timestep-shift strategy with identity weights frozen, ensuring smooth and coherent animation.
Extensive experiments demonstrate that FairyGen effectively generates personalized animated stories that are stylistically consistent, narratively coherent, and rich in natural motion.
In summary, our main contributions are as follows:
\begin{itemize}
  \item We present a novel story video generation framework that synthesizes stylistically consistent, narratively coherent, and temporally smooth animations from a single child-drawn character image.
  \item We propose a novel style propagation adapter that learns from character illustrations and generates background scenes in a compatible style while preserving character-specific visual and semantic features.
  \item We demonstrate that shifting diffusion timesteps in the image-to-video generation significantly enhances the model's ability to learn natural, fluid motions.
\end{itemize}

\section{Related Work}
\begin{figure*}[t]
    \centering
    \includegraphics[width=\textwidth]{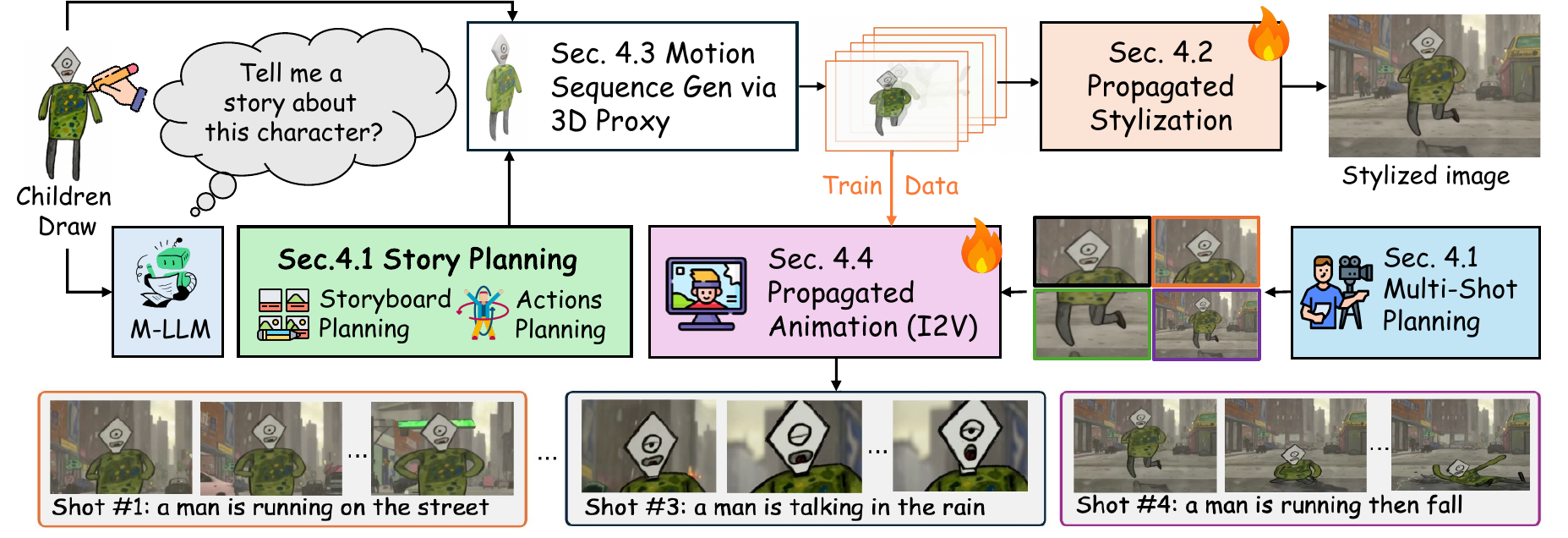}
    \vspace{-2em}
    \caption{\textbf{\textit{The pipeline of the whole FairyGen}}. }
    \label{fig:pipeline}
\end{figure*}
\noindent\textbf{\textit{Story Generation.}}
Generating the visual story from the text description contains many challenging problems, including the consistency of the overall style of the video, the consistency of the subject, \etc. Early works directly train models based on a well-prepared small dataset utilizing a GAN-based framework~\cite{storygan, li2022word} or transformer~\cite{chen2022character,maharana2022storydall} to directly generate the story video.
Recently, image and video diffusion models~\cite{ddpm, LDM} and the Large Language Models~\cite{chatgpt, qwen, deepseek-r1} bring the universal generative priors to this problem for planning, generation, and animation.
SEED-Story~\cite{seed-story} directly trains the Large Language Model for the consistency and long story visualization.
As for diffusion-based methods,
TaleCrafter~\cite{talecrafter} proposes a multi-stage framework to generate the visual story from the customized text-to-image model with a traditional camera movement, which is then extended by AutoStory~\cite{autostory} with more conditions.
StoryDiffusion~\cite{story-diffusion} proposes a customized multi-frame attention layer for consistent identity generation, similar to \cite{Liu_2024_CVPR}. StoryAgent~\cite{story-agent} involves the multiple stages of the LLM as the agent for storytelling, \etc. However, all these methods~\cite{tao2024storyimager, zhang2025storyweaver, su2023make,zhu2025cogcartoon} focus mainly on a custom identity pipeline for characters, then the video can be generated via the pre-trained video diffusion model. We argue that this kind of pipeline is still hard to customize the child-draw images with a similar corresponding style, and is difficult to generate a suitable motion utilizing the generation prior directly. Different, to avoid the complex movement of the foreground character, we involve a 3D proxy to solve the complex motion and the character consistency. 

\noindent\textbf{\textit{Customized Generation.}}
In the era of the large generative model, 
fine-tuning the text-to-image/video models~\cite{videocrafter1, videocrafter2, wanvideo, hunyuanvideo, yang2024cogvideox} for customizing usage is natural and practical. 
Our method also shares a similar inspiration with customization methods since we need to generate the customized background style and motions. For subject customization, 
early works~\cite{dreambooth, lora, Textual-Inversion} focus on single subject customization via additional parameter-efficient training, as well as multiple subjects~\cite{celeb-basis, customDiffusion}, and also for video~\cite{jiang2024videobooth}. As for both subject and motion customization.
DreamVideo~\cite{Dreamvideo}, CustomTTT~\cite{customttt}, MotionDirector~\cite{Motiondirector} trains different LoRAs~\cite{lora} for appearance and motion customization. DynamicConcept~\cite{dynamic-concept} trains a two-stage LoRAs~\cite{lora} for customized motion generation.
Stylization customization shares a similar idea behind the subject customization. For example, B-LoRA~\cite{B-LoRA} splits the subject and style via layer-specific learning. StyleDrop~\cite{styledrop} increases the stylization effects via iterative training. 
However, it is still unclear how to utilize these methods in our child-draw background generation with a given foreground character and in the image-to-video framework.

\section{Preliminarires}
\noindent\textit{\textbf{Latent Diffusion Model.}} Most of the current generation models are based on the latent diffusion model~\cite{sd}. Take the image diffusion model as an example, it contains a pre-trained VAE to encode/decode the image to the latent space. Then, the diffusion model aims to train a denosing network to remove the added single-step noise via the simple MSE loss. Formally, given the image $I$ and its corresponding latent $z = \mathcal{E}(I) $, we first add the $t$ step noise $\epsilon$ to the latent $z$ for $z_t$ where we train a denoising network to remove the added noise via:
\begin{equation}
    \mathcal{L} = || \epsilon -  \epsilon_\theta(z_t, c, t) ||_2,
\end{equation}
where $c$ is the condition signal, which is often the text features from the pre-trained text encoder~\cite{clip, t5}. After training, the image can be generated from noise via the multi-step sampling process and VAE decoding.

\noindent\textit{\textbf{LoRA for Customization.}} LoRA~\cite{lora} is first proposed for efficiently training the language model. Given the weight $W \in \mathbf{R}^{m \times n}$ of any pre-trained model, LoRA only updates the parameters of the two low-rank matrices $A \in \mathbf{R}^{m\times l} $ and $B \in \mathbf{R}^{l \times n}$, where $l < m$ and $ l < n $, to efficiently adapt the trained knowledge to the learned domain, which can be defined as: $ y = Wx + ABx $, where $x$ and $y$ is the input vector and new weighted vector, respectively. In the diffusion model, LoRA is a common customization technique to tune the weight of $\epsilon_\theta$ using customized datasets.

\section{Method}
Given a single child hand-drawn character image $I$ with a blank background, our goal is to generate a fully stylized, long cartoon video $V$ that unfolds as a continuous story composed of multiple distinct shots. The generated video should faithfully maintain character consistency while enabling complex motion, coherent scenes, and cinematic storytelling.
As illustrated in Fig.~\ref{fig:pipeline}, we propose a multi-stage pipeline to achieve this goal. 
First, we employ an MLLM~\cite{chatgpt} to infer a structured storyboard from the given character draft, forming the basis for temporal and spatial shot planning (Sec.~\ref{sec:story-planning}). Next, conditioned on the input stylized image, we synthesize style-consistent backgrounds using a customized style propagation module, which propagates the character’s aesthetic traits to the surrounding environment, introducing rhythmic pacing and diverse spatiotemporal cues crucial for cinematic storytelling~(Sec.~\ref{sec:style}). 
Then, we reconstruct a 3D proxy from the 2D character image, and derive physically plausible motion sequences by rigging and retargeting~(Sec.~\ref{sec:motiongen}). These motion sequences are then used as trainning data to finetune an MMDiT-based image-to-video diffusion model network, where we utilize a two-stage LoRA training scheme to disentangle spatial identity and temporal motion features~(Sec.~\ref{sec:motion}). 
Overall, this pipeline preserves character identity, boosts motion fidelity, and heightens narrative tension. We discuss each part in detail in the following section.

\subsection{Story and Shot Planning from a Single Character}
\label{sec:story-planning}
\begin{figure*}
    \centering
    \includegraphics[width=\textwidth]{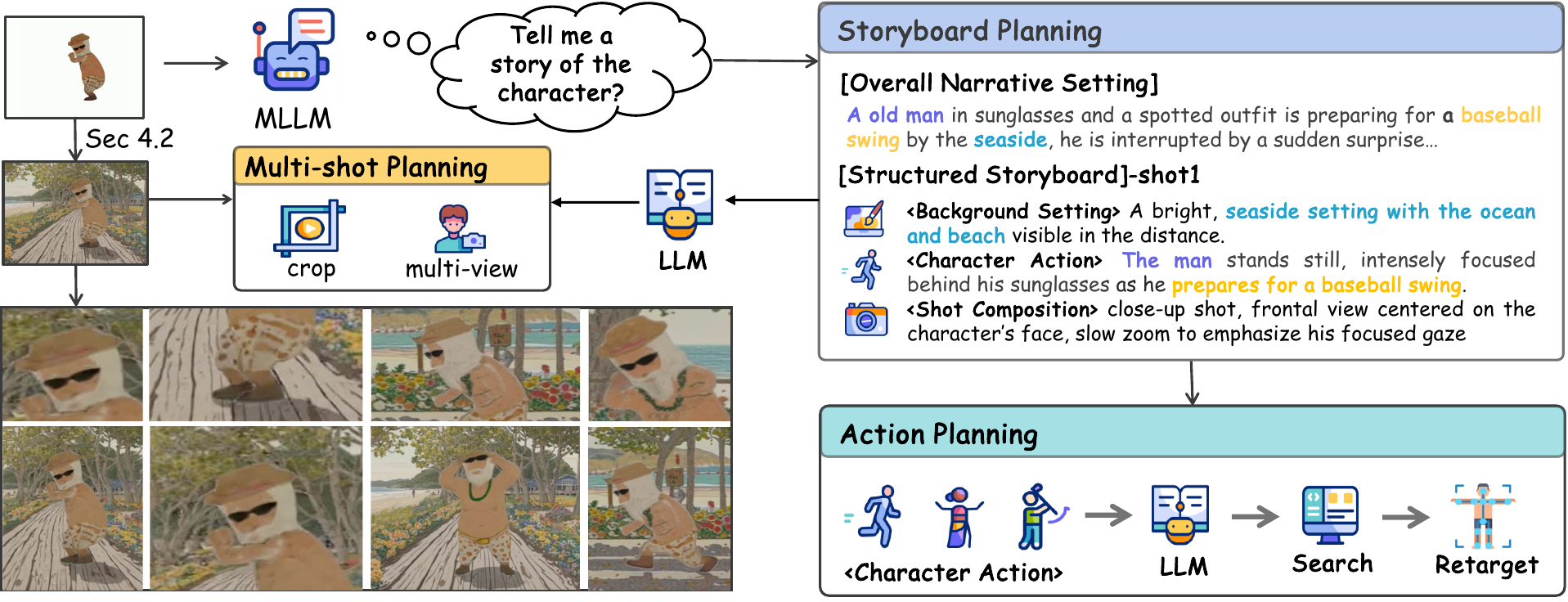}
    \caption{\textbf{\textit{The pipeline of the storyboard generation.}} We first plan the whole story using the M-LLM and build a storyboard containing the scenes, events, character action, background, and camera shots.  Then, we crop the stylized image using different camera shot and generate final shot images.}
    \label{fig:storyboard}
\end{figure*}

To enable narrative-driven video synthesis from a single character draft, we introduce a story and shot planning module that decomposes the narrative into cinematic shot descriptions. Unlike prior work that focuses on low-level frame interpolation, our approach defines a clear storyboard to guide motion synthesis, shot composition, and narrative pacing. By grounding animation in a storyboard, we ensure coherent spatial composition and temporal progression aligned with storytelling conventions.

As illustrated in Fig.~\ref{fig:storyboard}, our system begins with\textbf{ }storyboard planning, which organizes the narrative through a hierarchical structure: a global narrative overview and a detailed shot-level storyboard. The global narrative outlines the character's appearance, background context, and a high-level abstraction of the main event. Furthermore, the shot-level storyboard details the background, character action and camera configurations (\eg, shot type, perspective, focal region) for each shot. To operationalize the storyboard’s components, we introduce two modules: action planning and multi-shot planning. In action planning stage, action-related keywords are extracted using an LLM and then used to retrieve matching motions from a 3D animation platform. These motions are then adapted to the input character through rigging and retargeting. In the multi-shot planning stage, shot type and focal region descriptions guide the generation of bounding boxes, via an LLM, to crop synthesized backgrounds. Meanwhile, for multiview consistency, various character perspectives are rendered using a 3D proxy (see Sec.~\ref{sec:motiongen}), and view-conditioned synthesis is applied to ensure coherent background generation (see Sec.~\ref{sec:style}).

\subsection{Style-Consistent Scene Generation from Character}
\label{sec:style}
\begin{figure}[t]
    \centering
    \includegraphics[width=\columnwidth]{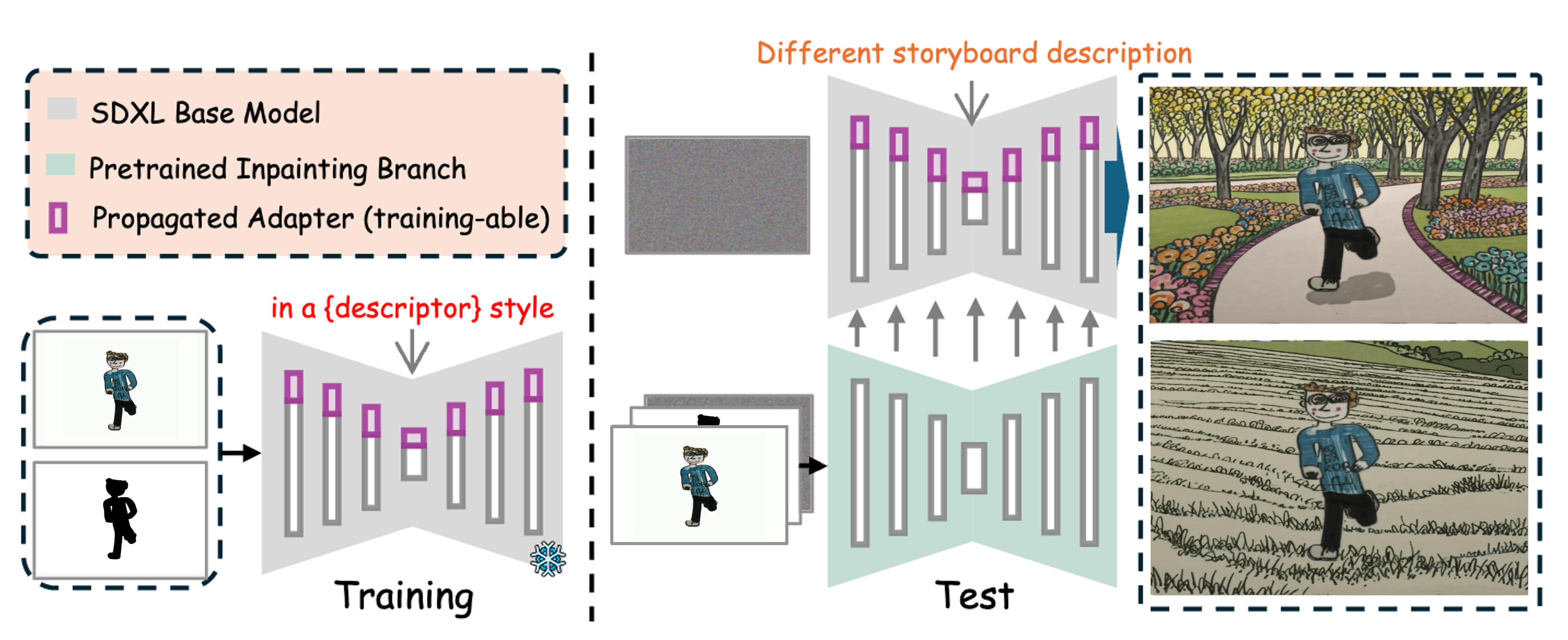}
    \caption{\textit{\textbf{Style Consistent Scene Generation}}.}
    \label{fig:style}
\end{figure}
A character image with no background is insufficient for expressive storytelling. To visually support the narrative, we aim to generate scenes that are both contextually aligned with the storyline and stylistically consistent with the hand-drawn aesthetic of the foreground character. This consistency is particularly crucial in cartoon story videos, where visual uniformity enhances continuity, immersion, and emotional resonance.
A key challenge is ensuring that the generated background faithfully reflects the artistic style, such as brushstroke texture, color palette, and line density, of the original hand-drawn character. Unlike traditional stylization approaches that transfer style from a reference background, our method propagates style from the character to the background, requiring the background to inherit the foreground’s visual attributes.

To achieve the goal, we start from a pre-trained text-to-image diffusion model, \ie, SDXL~\cite{sdxl}, and adapt it using a propagation-based customization strategy.  As shown in Fig.~\ref{fig:style}, in training, we customize only the foreground tokens to learn the artistic style, and for inference, we utilize the BrushNet~\cite{brushnet} adapter based on SDXLto inpaint the background using the learned style, which is propagated through the adapter. Notably, the adapter is applied only to background tokens during inference to maintain targeted stylization.
In detail, our propagated adapter is implemented as a low-rank adapter~\cite{lora, dora}. Specifically, we find that DoRA~\cite{dora} demonstrates better performance in capturing stylistic detail.

Formally, let $x$ donate the full image tokens, and $m$ be the binary foreground mask. The propagated adapter $PA(\cdot)$ updates the model as follows during training:
\begin{equation}
y = Wx + PA(x \cdot m),
\end{equation}
where $W$ is the original weights of SDXL, and $y$ is the customized feature output.

During inference, we select the background region for image stylization, which can be formulated as:
\begin{equation}
y = Wx + PA(x \cdot (1 - m)).
\end{equation}

In summary, our method learns the character’s visual style during training and selectively propagates it to the background during inference. This simple yet effective strategy ensures that the generated scenes remain stylistically unified with the character, supporting coherent and visually immersive animation.

\subsection{Character Video Sequence Generation via 3D Proxy}
\label{sec:motiongen}
Previous work learns to generate id-consistent video using customized image generation methods~\cite{dreambooth}, and then generates video using image-to-video diffusion models for storytelling \cite{talecrafter}. However, restricted by the generative prior of the image-to-video diffusion model, generating id-consistent and coherent motion is inherently challenging, since the foreground human motion is very complex. 

Differently, we draw inspiration from traditional computer graphics pipelines, which enable fine-grained control over character motion through intermediate 3D representations. Specifically, following \textit{DrawingSpinUp}~\cite{DrawingSpinUp}, we adopt a 3D proxy-based motion modeling approach that reconstructs the underlying 3D geometry of the character from a single 2D sketch. This proxy enables us to apply skeleton-based rigging and motion retargeting techniques, allowing for the transfer of complex motion sequences onto the character while preserving structural fidelity and visual consistency. 
By incorporating this explicit motion structure, we provide a robust foundation for generating animation semantically meaningful and visually coherent animation sequences. More details can be found in the original paper.

\subsection{Shots Animation via Motion Customization}
\label{sec:motion}
\begin{figure}[h]
    \centering
    \includegraphics[width=\columnwidth]{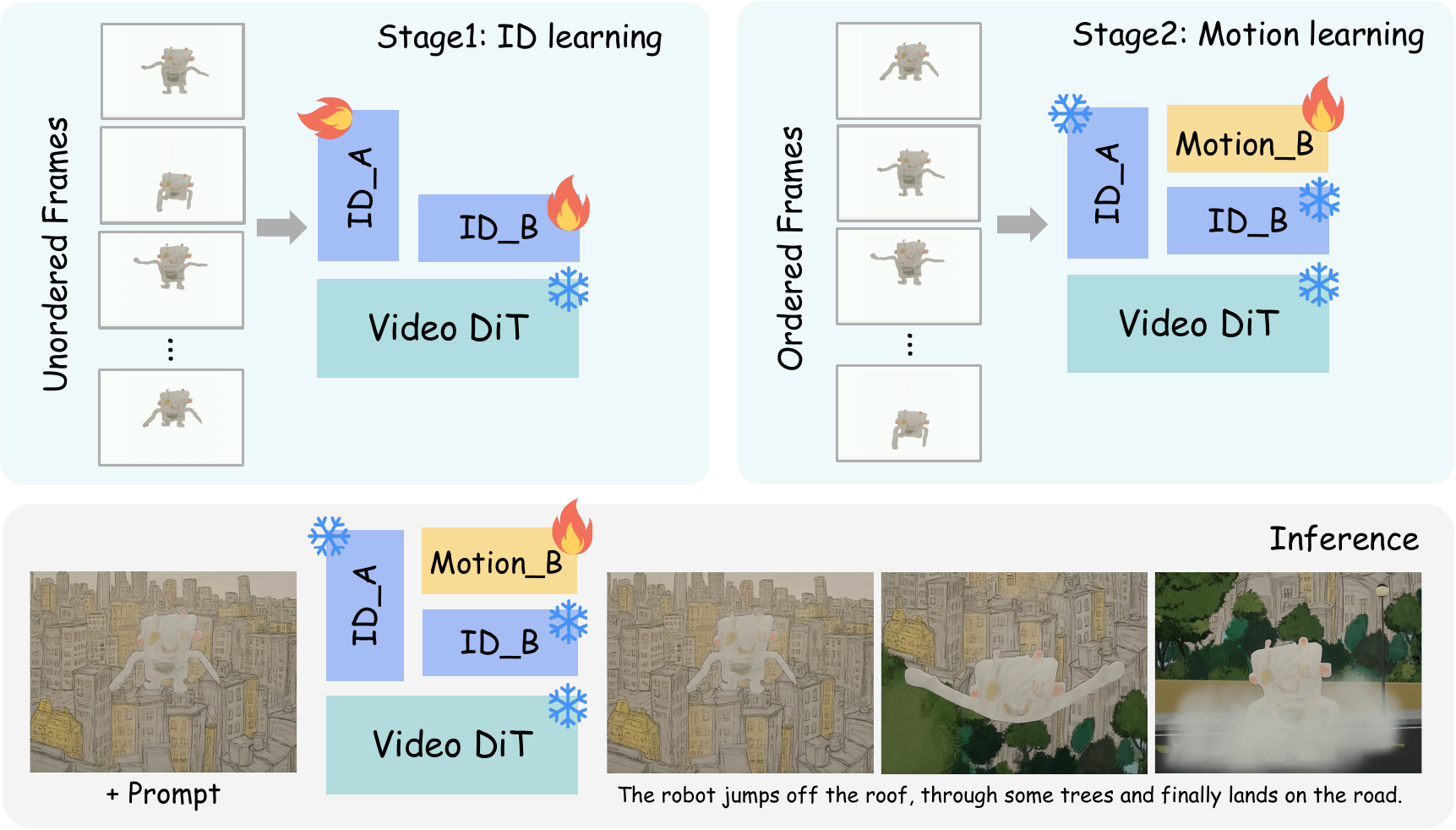}
    \caption{\textbf{\textit{Two-stage motion train stratage.}} We first use unorded frames to learn character spatial features without temporal bias. Then, with the identity LoRA frozen, motion residuals are learned from sequential video frames.}
    \label{fig:motion_train}
\end{figure}

To generate animatable video shots from background-composited frames, we leverage an image-to-video diffusion model. However, existing image-to-video diffusion models \cite{svd, wanvideo, hunyuanvideo} struggle to generate complex motion for stylized or anthropomorphic characters, resulting in identity inconsistency and temporal flickering. Moreover, the ControlNet~\cite{controlnet}-like video control models~(\eg, pose-guided and depth-guided) exhibit limited generalization to non-human characters and often produce unnatural or scene-disconnected motion due to overly rigid constraints.
Differently, we leverage character motion sequences derived from 3D reconstruction as train data to finetune the video diffusion model. Our main challenge is achieving shot-level animation, where only specific body parts (\eg, head or legs) are animated. Direct training under such partial-motion often fails to maintain appearance consistency across frames. Additionally, existing video diffusion models require extensive training iterations to learn complex motion patterns, even for a single sequence.
To address these challenges, we adopt a two-stage training strategy inspired by DynamicConcept~\cite{dynamic-concept}, , explicitly disentangling spatial appearance learning from temporal motion learning, as illustrated in Fig.~\ref{fig:motion_train}. 
In the first stage, the model is trained on temporally shuffled frames to learn identity features without temporal correlations. In the second stage, the identity adapter is frozen, and a separate motion adapter is introduced. The model is then trained on temporally ordered frames using a novel timestep-shift strategy to effectively capture temporal dynamics.
Instead of replicating motions exactly, our stategy learns to generate motion sequences that adapt to diverse backgrounds and narative contexts. During inference, the learned motion is composited with background-composited scenes to produce coherent, stylized animations.
Let $W$ donate the base weights of the video diffusion model, and let $A_{\text{id}}, B_{\text{id}}$ be the low-rank identity adapter matrices \ie, LoRA~\cite{lora}. The identity-adapted features are computed as:
\begin{equation}
y = Wx + A_{\text{id}} B_{\text{id}} x,
\end{equation}
where $x$ is the input feature. To prevent the model from inadvertently learning temporal patterns during identity training, we apply dropout to $B_{\text{id}}$:
$B_{\text{id}} = B_{\text{id}} \odot M_p$, 
where $M_p$ is a binary mask with dropout probability $p$. 
In the second stage, we fix the identity adapter and introduce a motion-specific adapter $B_{\text{motion}}$, applied to sequential frames. Motion is modeled as a residual deformation on top of the identity representation:
\begin{equation}
y = Wx + A_{\text{id}} B_{\text{id}}x + A_{\text{id}} B_{\text{motion}}x.
\end{equation}
Dropout is also applied to $B_\text{motion}$ to stabilize training and prevent overfitting, ensuring that $A_{\text{id}}$ remains a stable shared basis across both training stages.

While this two-stage training effectively separates motion from appearance, we further improve motion modeling through a novel timestep-shift sampling strategy, which we identify as critical for capturing realistic, coherent character dynamics in image-to-video motion customization. 
Standard diffusion training samples timesteps $t \in \{1, \dots, T\}$ uniformly, emphasizing clean and noisy frames equally. However, we hypothesize that biasing training toward noisier timesteps—later in the diffusion process—forces the model to rely on global structure rather than low-level pixel cues. We implement this bias using Gaussian sampling followed by a sigmoid transformation:
\begin{equation}
t = \sigma(z) = \frac{1}{1 + e^{-z}}, \quad z \sim \mathcal{N}(\mu, \sigma^2)
\end{equation}
where $\mu$ controls the sampling bias and $\sigma$ the variance. By setting $\mu$ closer to $T$, we construct a late-bias sampling distribution that increases the probability of sampling high-noise training steps. This late-biased sampling encourages the model to learn robust motion representations under challenging conditions. Empirically, we observe that this strategy produces smoother and more temporally consistent motion trajectories, especially for long sequences with complex character interactions.

\begin{figure*}
    \centering
    \includegraphics[width=1.0\linewidth]{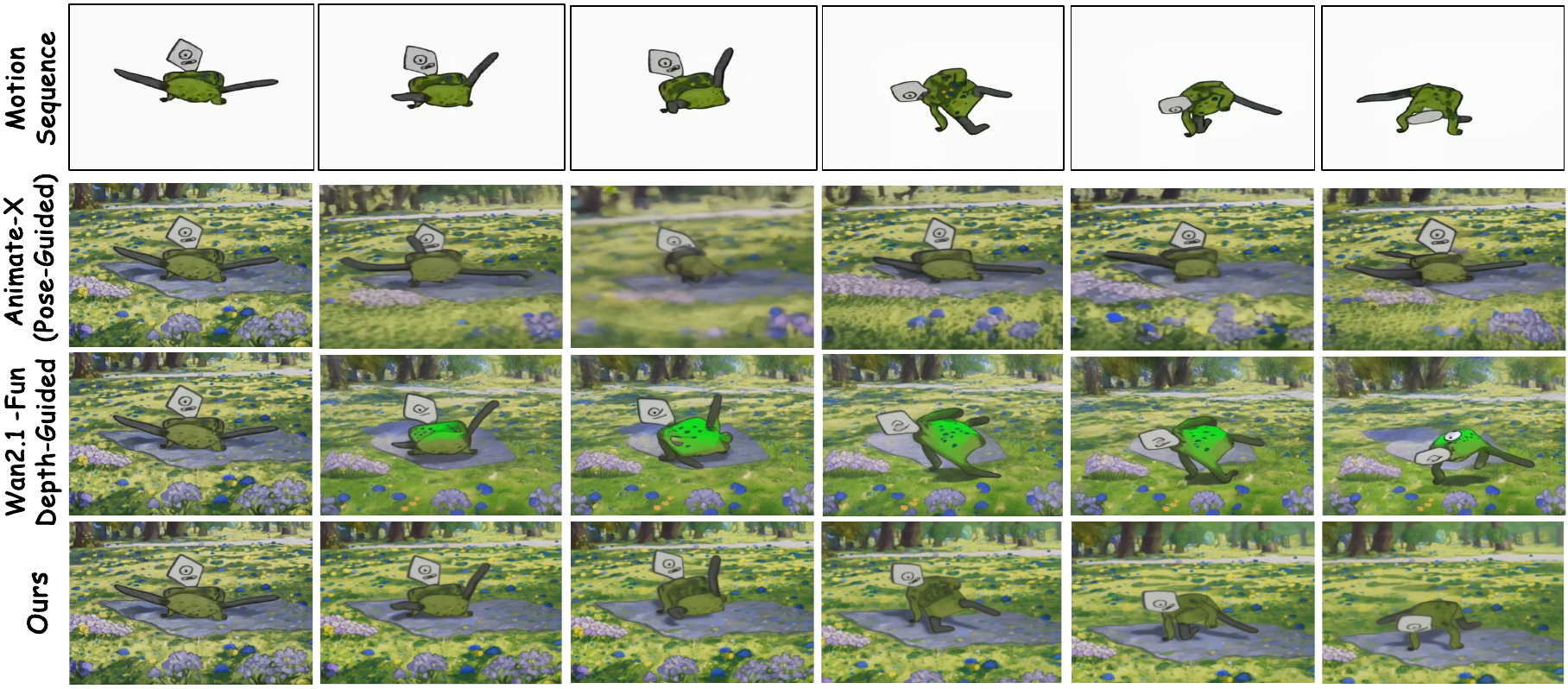}
    \caption{ \textit{\textbf{Compare with Motion Customization}.} We compare the proposed motion customization method with the depth-guided image-to-video method using Wan2.1~\cite{wanvideo}, pose-guided image-to-video character animation method, \ie, Animate-X~\cite{tan2024AnimateX}, the proposed method shows very similar results to the original motion sequence with this complex motion. }
    \label{fig:compare_motion}
\end{figure*}

\section{Experiment}
\subsection{Implementation Details}

For the experiment and comparison with other methods, we utilize the AnimatedDrawings Dataset~\cite{animated-draw} as our child-drawn characters. We generate 24 images of different styles and 12 videos with different motions for style and motion comparison, respectively.  All the experiments are based on one NVIDIA L20 GPU, it takes 120 minutes to learn the style and 180 minutes for motion customization.
For stylization, instead of relying on artificial identifiers (\eg, ``a [v] style”) as in DreamBooth, we find that using \textit{descriptive language prompts} (\eg, ``a childlike whimsical style”) yields better alignment with fine-grained stylistic attributes such as texture, brushstroke direction, and line quality. To automate this, we employ GPT-4~\cite{chatgpt} to generate textual descriptions of the reference image's style, which are then used as part of the training prompt similar to StyleDrop~\cite{styledrop}. After training, we can use text to generate the background style and motion.

For style evaluation, we calculate the CLIP~\cite{clip} distance of the generated image and the source image as a style alignment score, where we also calculate the CLIP distance of the generated image and the corresponding CLIP text features. As for motion evaluation, we choose two metrics from the VBench~\cite{vbench, evalcrafter}, including the motion smoothness and the subject consistency.

\begin{figure}[b]
    \centering
    \includegraphics[width=1.0\linewidth]{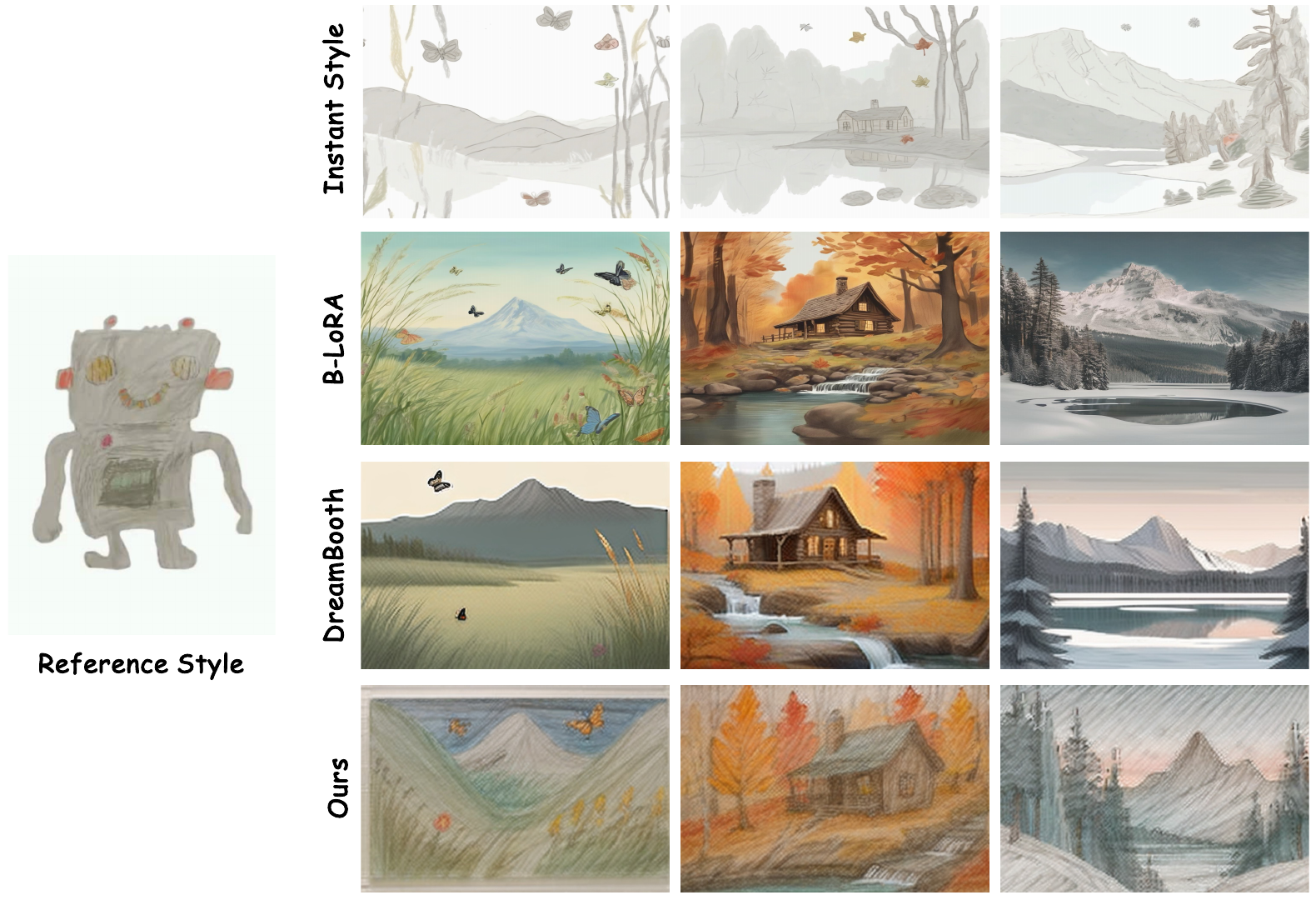}
    \vspace{-2em}
    \caption{\textbf{\textit{Compare with Stylization Methods.}} We compare our method with different stylization methods on stylization customization.
    }
    \label{fig:compare_style}
\end{figure}

\subsection{Comparison with Other Methods}
\begin{figure*}
    \centering
    \includegraphics[width=1.0\linewidth]{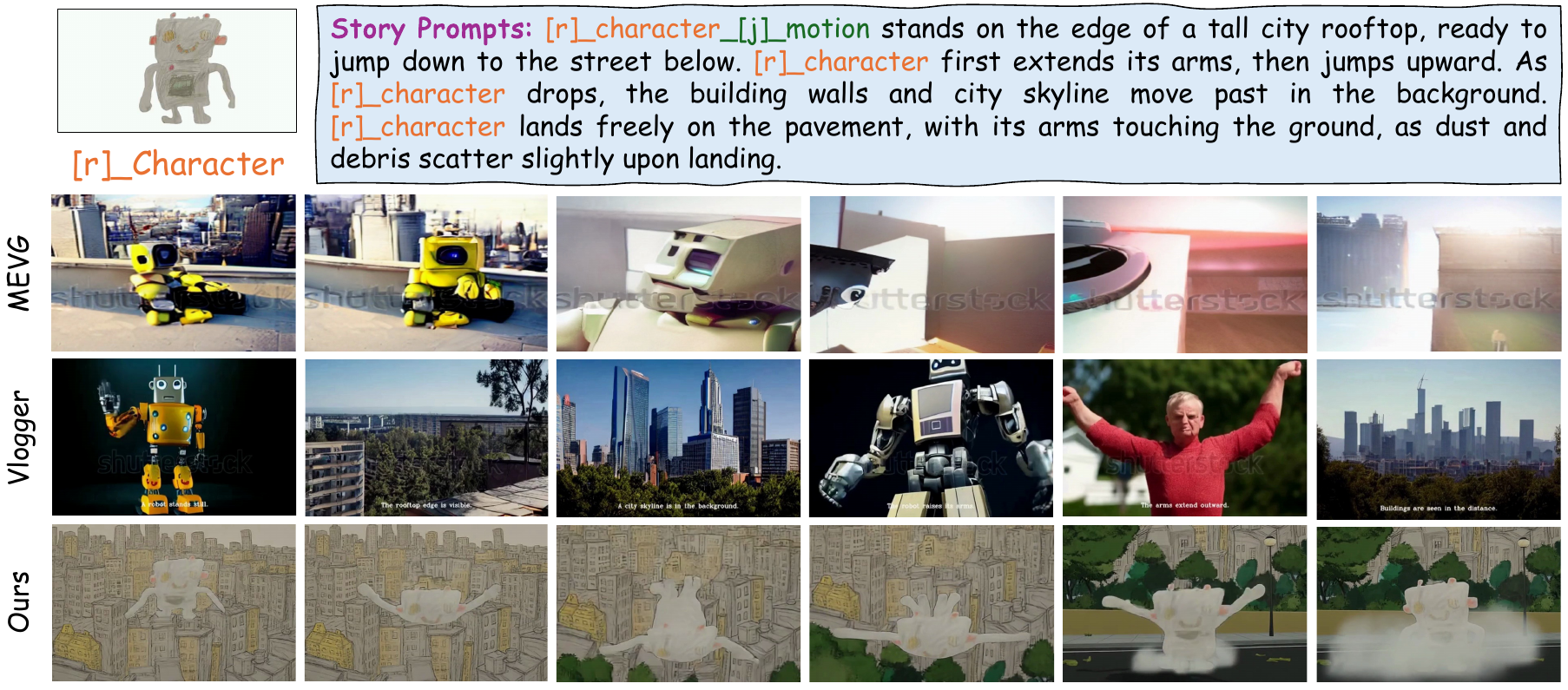}
    \vspace{-2em}
    \caption{\textbf{\textit{Comparsion on Multi-Event Video Generation.}} Our method splits the foreground and the background modelling, which is easy for longer and multi-event video generation. Here, we use the same story prompt to generate the video, where the proposed method shows much consistent results as well as the text description. }
    \label{fig:compre_story}
\end{figure*}

\begin{figure*}
    \centering
    \includegraphics[width=1.0\linewidth]{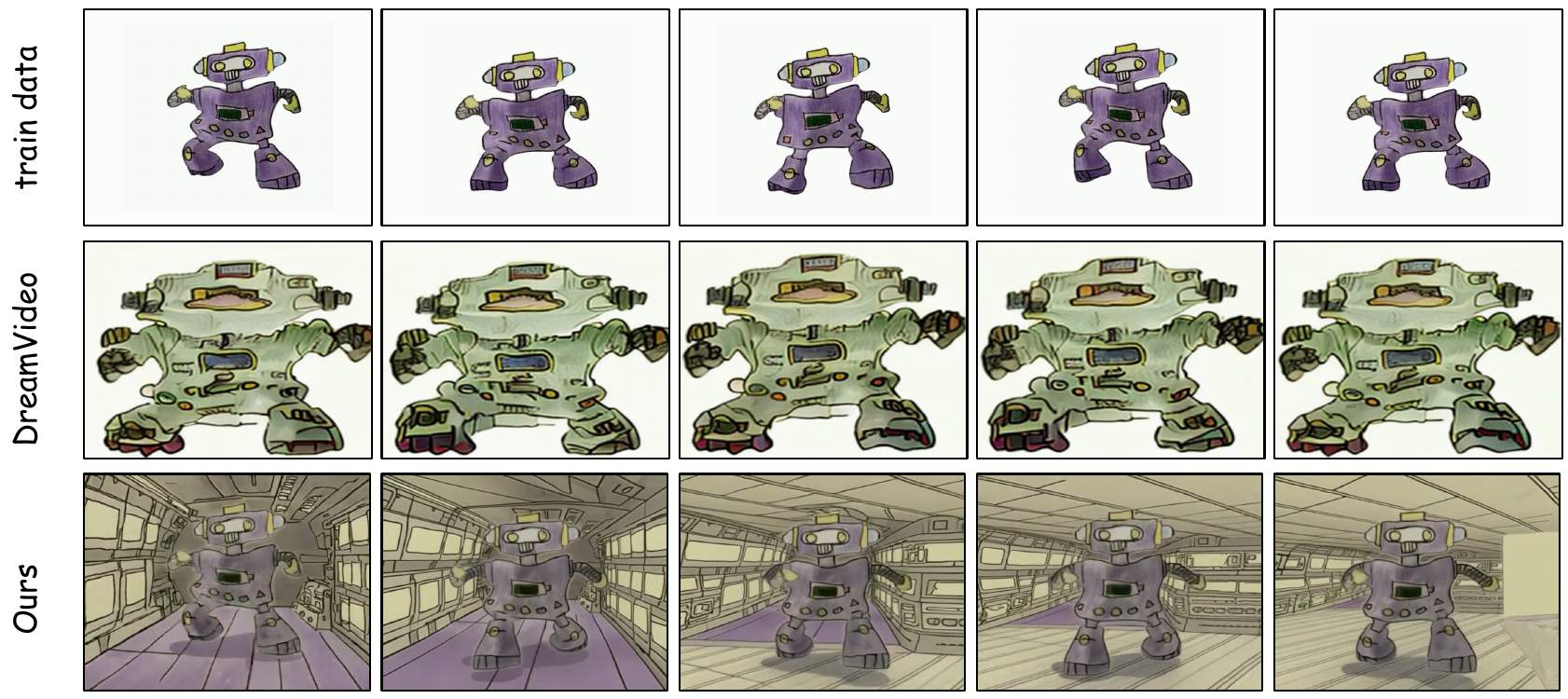}
    \vspace{-2em}
    \caption{\textbf{\textit{Comparison on the appearance and motion customization method.}} We compare our method with state-of-the-art appearance and motion customization method, \ie, DreamVideo~\cite{Dreamvideo}, the proposed method shows significantly better results considering the stylization, motions, and overall quality. }
    \label{fig:compre_id_motion}
\end{figure*}

Since there are no previous baselines for this task, we first compare our method with multi-event video generation from text description, \ie, MEVG~\cite{mevg} and Vlogger~\cite{vlogger},  as well as the state-of-the-art few-shot subject-driven video generation method, \ie, DreamVideo~\cite{Dreamvideo}. As shown in Fig.~\ref{fig:compre_story} and Fig.~\ref{fig:compre_id_motion}, the proposed method shows better results in maintaining character appearance consistency, motion smoothness, and style preservation. In contrast, multi-event generation models struggle to produce coherent narratives and consistent visual styles, while subject-driven methods fail to robustly preserve identity and motion, and often produce unrealistic or inconsistent backgrounds. By leveraging a 3D proxy as motion prior and a customized style propagation adapter, our method achieves more coherent, stylistically faithful, and temporally smooth video generation results.
 
\begin{table}[t]
    \centering
    \begin{tabular}{lcccc}
    \toprule
    & \multicolumn{2}{c}{Numerical Comparsion} & \multicolumn{2}{c}{User Study} \\
    \cmidrule(lr){2-3} \cmidrule(lr){4-5}
 Methods  & Style  & Text   &  Style   & Visual \\ 
          & Align  & Align   &  Quality & Quality \\ \midrule
       B-LoRA    &   0.5060  &  \textbf{0.2829}      &  0.0267        &  \textbf{0.3429}  \\
       Instant Style  &  0.5468   &  0.2368    &  0.3403       &  0.0517  \\
       DreamBooth    &   0.6371  &   0.2819  &   0.0965        &   0.2803  \\
       Ours          &  \textbf{0.6580}   &   0.2702   &   \textbf{0.5365}      &   0.3251  \\ \bottomrule
    \end{tabular}
    \caption{\textit{\textbf{Style Comparsion}} with other methods.}
    \vspace{-2em}
    \label{tab:style}
\end{table}

\begin{table}[t]
    \centering
    \begin{tabular}{lccccc}
        \toprule
    & \multicolumn{2}{c}{Numerical Comparsion} & \multicolumn{2}{c}{User Study} \\
    \cmidrule(lr){2-3} \cmidrule(lr){4-5}
    Methods   & Motion & Subject & Motion  &  Visual \\ 
         & Smooth. & Consist. &  Realness &  Quality \\ \midrule
       Animate-X  & 0.974 &0.908 & 0.106 & 0.023\\
       Wan2.1-Fun  & 0.977 &0.842 & 0.114 & 0.106\\ 
       Ours  & \textbf{0.987} & \textbf{0.955} & \textbf{0.780} & \textbf{0.871}\\ 
       \bottomrule
    \end{tabular}
    \caption{\textit{\textbf{Motion Comparsion.}} with other methods.}
    \label{tab:motion}
\end{table}

\begin{figure}[t]
    \centering
    \includegraphics[width=1.0\linewidth]{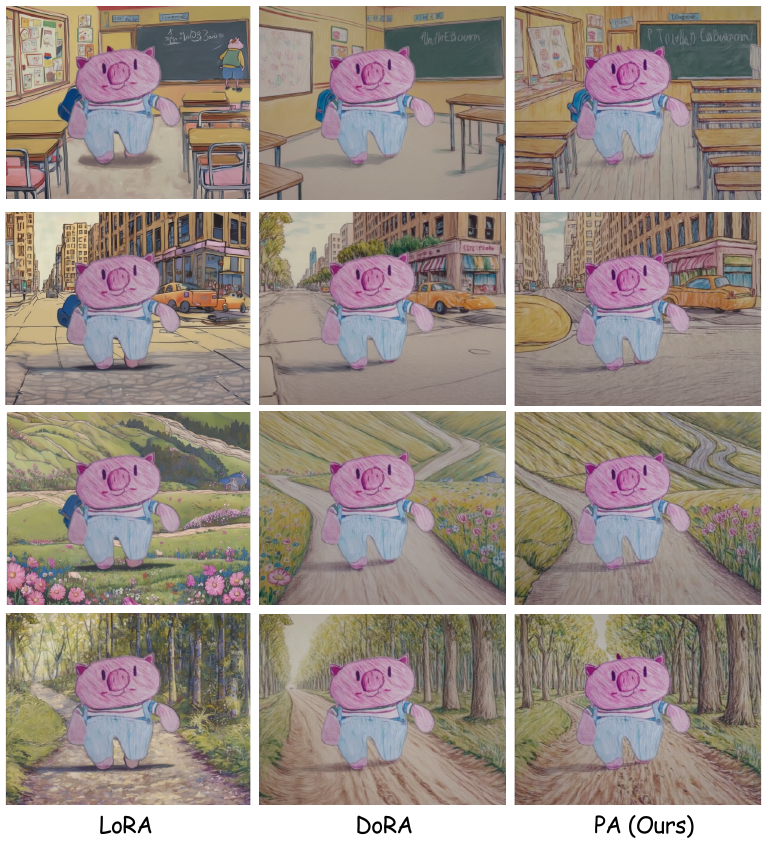}
    \vspace{-2em}
    \caption{ \textbf{\textit{Ablation Study on Style Customization.}} Compared with the baseline LoRA~\cite{lora} and DoRA~\cite{dora}, the proposed method can successfully propagate the foreground style to the background with different prompts. Best viewed with zoom in.}
    \label{fig:ablation_style}
\end{figure}

In addition to visual comparison, we also perform detailed quantitative evaluations. For style comparison, we assess both the style alignment and the relevance to the accompanying text description. As shown in Tab~\ref{tab:style}, our method outperforms previous approaches in both subjective and objective stylization metrics.
As for the quality of motion,  we compare against two video character animation methods: the pose-guided method Animate-X, which uses human motion videos as reference for accurate keypoint detection, and a depth-guided method that utilizes depth sequences extracted from 3D-reconstructed character motion sequences. We compare the motion quality with the previous baselines in Tab.~\ref{tab:motion}, where the proposed method shows significantly better results. 

We further conduct subject experiments on both stylized images and generated videos to evaluate the effectiveness of the proposed method.  As shown in Tab.~\ref{tab:motion} and Tab.~\ref{tab:style}, we invite 24 users to evaluate 24 stylized image sets, each set contains 4 different methods, and needs to be evaluated from two aspects. As for motion, we utilize 12 video sets using 3 different methods, and the users need to be evaluated from two aspects. Finally, we obtain 3360 opinions. 
As shown in tables, the users consistently prefer our method in terms of style alignment, motion realism, and visual coherence. As shown in Tab.~\ref{tab:style}, our method achieves the highest score in style similarity, surpassing B-LoRA, InstantStyle, and DreamBooth. While our visual impression score (0.3251) is slightly lower than B-LoRA (0.3429), we attribute this to B-LoRA's photorealistic output, which may be perceived as more visually appealing than child-style cartoon imagery. For video results (Table~\ref{tab:motion}), our method demonstrates clear advantages over existing approaches, highlighting its ability to produce temporally consistent, stylistically faithful animations.

\subsection{Ablation Studies}
\subsubsection{Propagation Style Adapter} We first evaluate the effectiveness of the proposed style propagation adapter. As shown in Fig.~\ref{fig:ablation_style}, given the image of the foreground character, the DoRA customization method shows a better stylization propagation result than the original LoRA method. Besides, the proposed propagation adapter can successfully learn the style better with more detailed streak preservation on the background.

\subsubsection{Motion Adapter} 
\begin{figure}[t]
    \centering
    \includegraphics[width=1.0\linewidth]{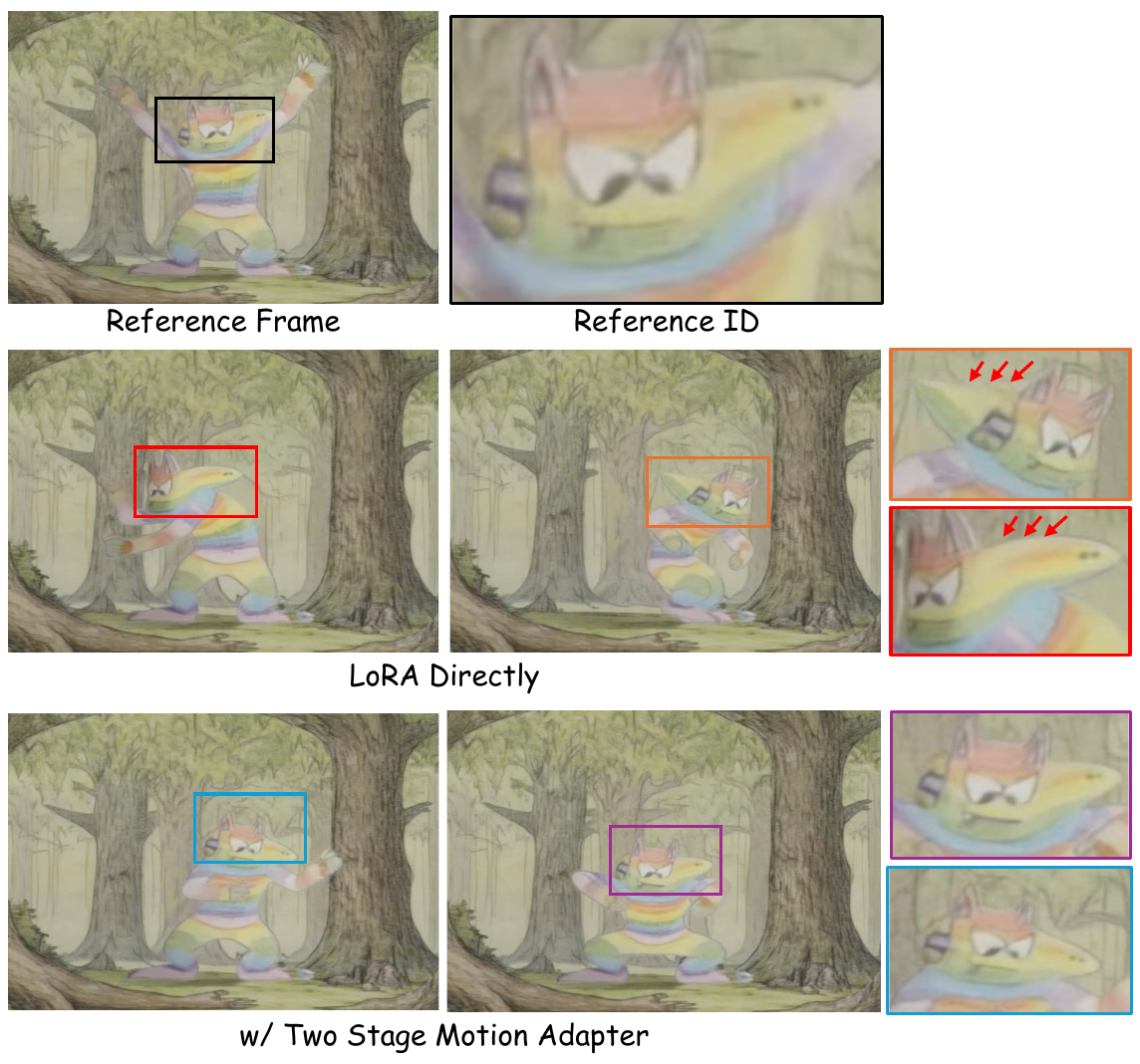}
    \vspace{-2em}
    \caption{\textbf{\textit{Ablation on two-stage Motion Adapter.}} We ablated the two-stage adapters in our proposed motion customization in image-to-video generation. Here, the first stage of training improves the identity similarity.}
    \vspace{-2em}
    \label{fig:ablation-two-stage}
\end{figure}

Our proposed Motion Adapter utilizes a two-stage customization strategy for motion customization. Here, we give the effectiveness of each stage. As shown in Fig.~\ref{fig:ablation-two-stage}, directly training the LoRA on the image-to-video diffusion model causes unnatural identity generation, while the two-stage motion adapter achieves better performance.

\begin{figure*}
    \centering
    \includegraphics[width=1.0\linewidth]{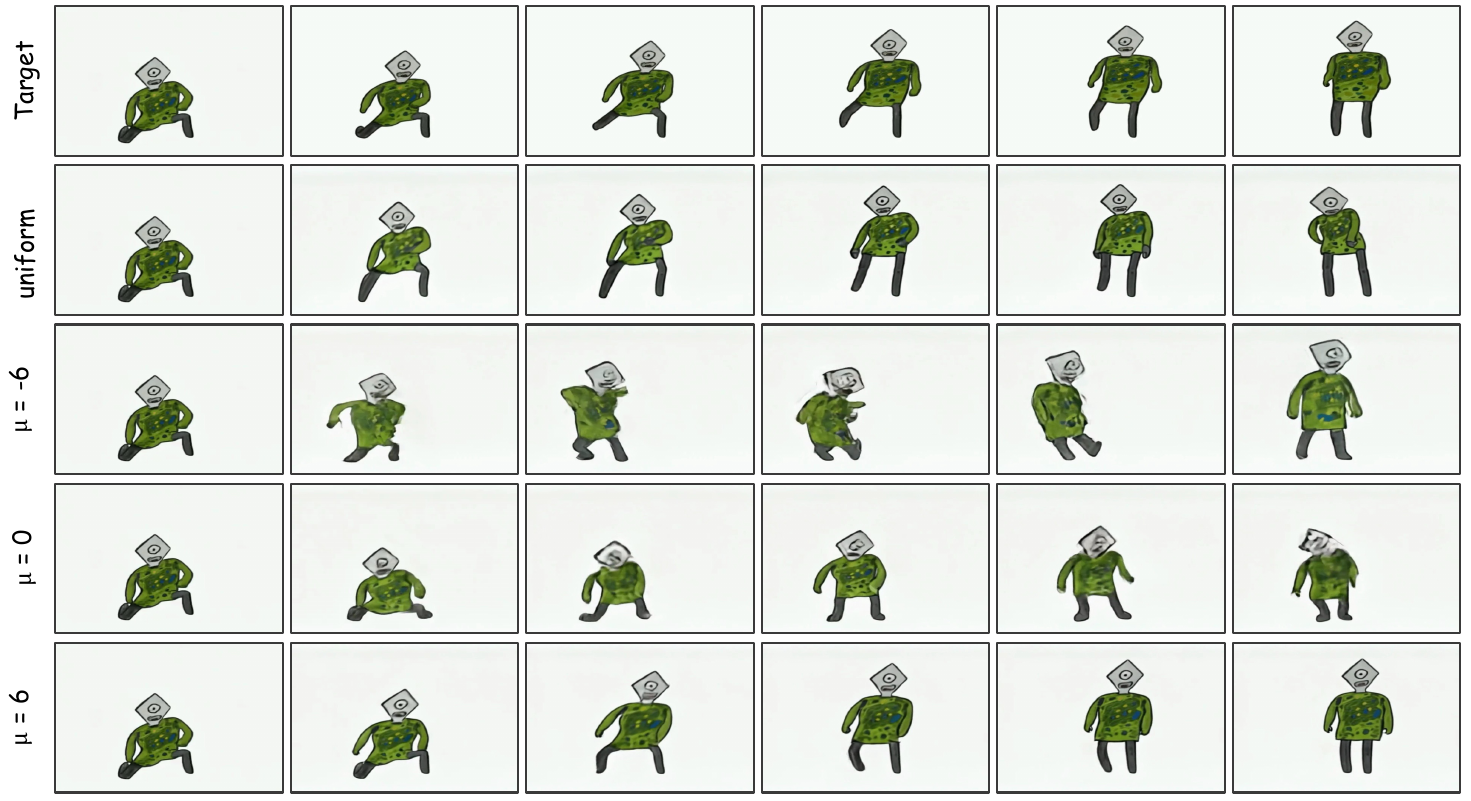}
    \vspace{-2em}
    \caption{\textbf{\textit{Ablation Study on timestep shift.}} The proposed timestep shift strategy in the motion customization can learn to represent the motion better. }
    \label{fig:ablation_motion_factor}
\end{figure*}

\subsubsection{Time-Shift for Motion Learning} One of our key techniques is to utilize the timestep shift in the step sampling to obtain better motion. We show the results of the different sampling steps in Fig.~\ref{fig:ablation_motion_factor}. As shown in the figure, directly using uniform sampling might not learn a similar motion to the original sample, while $\mu = 6 $ provides better results than uniform sampling and other hyperparameters.

\subsection{Limitation and Future Work}
\begin{figure}[h]
    \centering
    \includegraphics[width=\linewidth]{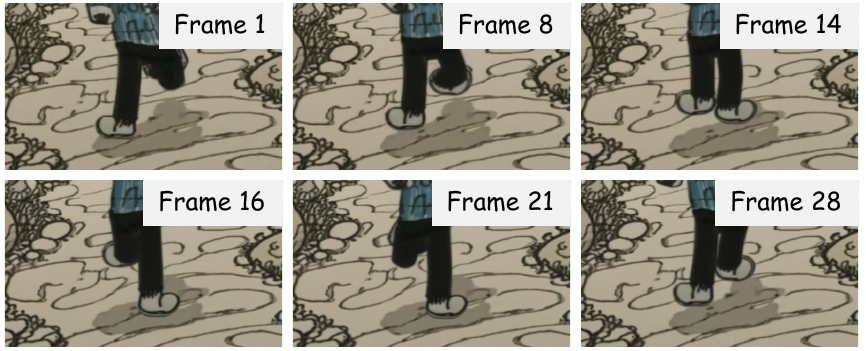}
    \vspace{-2em}
    \caption{\textbf{\textit{Limitation.}} Due to the uncontrollable generative prior of the video diffusion model, the proposed method might only generate a still background with the animated foreground motion~(\eg, running).}
    \label{fig:limitation}
\end{figure}

We only show the results of the single character. However, it is easy for our method to be extended into multiple subjects with multiple 3D proxies. 
The foreground character~(or animal) may not always be correctly reconstructed by the 3D proxy, we believe the more advanced rigging method~\cite{unirig} will help us to generate the motion of the foreground better. 
The generative prior of the video diffusion model may not always accurately generate a stable and animable background. 
As shown in Fig.~\ref{fig:limitation}, the proposed method generates a still background image. We will try different image-to-video diffusion models~\cite{hunyuanvideo} and include more camera motion~\cite{recammaster} to improve the motion realism.

\section{Conclusion}
We present FairyGen, a novel framework for story visualization that decomposes the narrative into foreground character motion and environmental dynamics, enabling layered modeling with unified style and movement. We first plan the storyline using a multi-modal large language model (M-LLM), followed by a style propagation adapter for generating stylized backgrounds consistent with the narrative context. To customize motion, we introduce motion-aware adapters and a timestep-sampling strategy for flexible control over character dynamics. Compared with several baselines, our method achieves high-quality results in stylized background generation and motion customization, demonstrating superior adaptability and visual coherence.


\bibliographystyle{ACM-Reference-Format}
\bibliography{sample-bibliography}


\end{document}